\documentclass[prd,superscriptaddress,twocolumn,nofootinbib,showpacs,preprintnumbers,floatfix]{revtex4}

\usepackage{mathrsfs}
\usepackage{amsfonts,amssymb,amsmath}
\usepackage{graphicx,epsfig}
\usepackage{multirow}
\usepackage{verbatim}
\usepackage{bbold}
\usepackage{units}

\def\fsu5{$\cal{F}$-$SU(5)$}
\def\bfsu5{$\boldsymbol{\mathcal{F}}$-$\boldsymbol{SU(5)}$}

\def\m1half{$M_{1/2}$}
\def\m3half{$M_{3/2}$}
\def\m32{$M_{32}$}

\def\fb{${\rm fb}^{-1}$~}

\def\mt2{$M_{T2}$}
\def\x2{$\chi^2$}
\def\2b{$M_{T2}b$}

\def\bs0{$B_S^0 \rightarrow \mu^+ \mu^-$}

\begin{document}

\title{Correlated Event Excesses in LHC SUSY Searches at 7 \& 8 TeV: \\ New Physics or Conspiring Noise?}

\author{Tianjun Li}

\affiliation{State Key Laboratory of Theoretical Physics and Kavli Institute for Theoretical Physics China (KITPC),
Institute of Theoretical Physics, Chinese Academy of Sciences, Beijing 100190, P. R. China}

\affiliation{George P. and Cynthia W. Mitchell Institute for Fundamental Physics and Astronomy, Texas A$\&$M University, College Station, TX 77843, USA}

\author{James A. Maxin}

\affiliation{George P. and Cynthia W. Mitchell Institute for Fundamental Physics and Astronomy, Texas A$\&$M University, College Station, TX 77843, USA}

\affiliation{Department of Physics and Astronomy, Ball State University, Muncie, IN 47306 USA}

\author{Dimitri V. Nanopoulos}

\affiliation{George P. and Cynthia W. Mitchell Institute for Fundamental Physics and Astronomy, Texas A$\&$M University, College Station, TX 77843, USA}

\affiliation{Astroparticle Physics Group, Houston Advanced Research Center (HARC), Mitchell Campus, Woodlands, TX 77381, USA}

\affiliation{Academy of Athens, Division of Natural Sciences, 28 Panepistimiou Avenue, Athens 10679, Greece}

\author{Joel W. Walker}

\affiliation{Department of Physics, Sam Houston State University, Huntsville, TX 77341, USA}


\begin{abstract}

We examine the ATLAS and CMS 7 \& 8 TeV multijet supersymmetry (SUSY) searches requiring the incidence of a single lepton in the framework of the supersymmetric grand unified model No-Scale Flipped $SU(5)$ with extra vector-like flippon multiplets derived from F-Theory, or \fsu5 for short. Investigated are five multijet+lepton SUSY searches: 4.7 \fb ATLAS 7 TeV gluino and light stop searches, as well as 13 \fb ATLAS and 9.7 \fb CMS 8 TeV light stop searches. Most significantly, all five leptonic SUSY searches represent statistically independent data samples. Findings show that all five orthogonal sets of leptonic LHC observations give a lower bound to the gaugino mass scale at $M_{1/2} \ge 680$ GeV, with all the current best fits correlating within a narrow region. Furthermore, eight statistically independent LHC SUSY search regions (leptonic + all-hadronic) accessible to the No-Scale \fsu5 model space intersect with all the currently operating beyond the Standard Model experiments within the range of $M_{1/2} = 680-850$ GeV, with the upper bound established by the lower experimental limit of the anomalous magnetic moment ${\rm (g_{\mu}-2)/2}$ of the muon. We emphasize that this region of the \fsu5 model space may not be fully probed by leptonic SUSY searches at the LHC until the 13 TeV LHC energizes in 2015. Additionally, we describe an efficient technique for the effective statistical disentanglement of searches sensitive to mutually overlapping event spaces.

\end{abstract}



\preprint{ACT-2-13, MIFPA-13-09}

\maketitle


\section{Introduction}

The LHC has accumulated 5 \fb of integrated luminosity at a beam collision energy of $\sqrt{s} = 7$ TeV in 2010--11 in the search for supersymmetry (SUSY), and an additional 20 \fb of data has been recorded in 2012 at $\sqrt{s} = 8$ TeV prior to the long shutdown in 2013--14 and energizing of the 13 TeV LHC in 2015. Much attention has shifted to the data collected at 8 TeV that is under analysis by the ATLAS and CMS Collaborations at present, and as we shall illuminate here, accumulating statistics manifest an intriguing profile of excess events that we believe could perhaps represent significant underlying correlations in the data observations for one prominent supersymmetric framework. Over and above this curious LHC data production, we further wish to elucidate interesting correlations of the 7 TeV results with this 8 TeV data set. If the peculiar relationships to be examined in-depth in this work are genuine and not the result of several parallel statistical fluctuations, then early data acquisition of 13 TeV collision statistics in 2015 should bear the first convincing evidence.

We have comprehensively studied the phenomenology and LHC collider analyses in the framework of the model No-Scale flipped $SU(5)$ with extra vector-like flippon multiplets, dubbed No-Scale \fsu5~\cite{Li:2010ws,Li:2010mi,Li:2010uu,Li:2011dw,Li:2011hr,Maxin:2011hy,Li:2011xu,Li:2011in,
Li:2011gh,Li:2011rp,Li:2011fu,Li:2011xg,Li:2011ex,Li:2011av,Li:2011ab,Li:2012hm,Li:2012tr,Li:2012ix,Li:2012yd,Li:2012qv,Li:2012jf,Li:2012mr}. While much of the focus of our previous work centered on the ATLAS designed SUSY searches~\cite{ATLAS-CONF-2012-033,ATLAS-CONF-2012-037,ATLAS-CONF-2012-103,ATLAS-CONF-2012-109} applying a large data cut on jet $p_T$, our originally suggested and preferred methodology attempted rather to preserve soft hadronic jets by establishing a lower $p_T$ threshold around 20 GeV~\cite{Li:2011hr,Maxin:2011hy}. We have shown that the flattening of the renormalization group equation (RGE) running of the $SU(3)$ gauge coupling from the GUT scale down to the flippon mass scale in No-Scale \fsu5 lowers the gluino mass below all the squarks except the light stop~\cite{Li:2010ws}. This then engenders a singular gluino decay channel to the light stop via $\widetilde{g} \to \widetilde{t}_1 \overline{t}$, most probably resulting in a 4-top final state for gluino pair-production events $\widetilde{g} + \widetilde{g} \to \widetilde{t}_1 \overline{t} + \widetilde{t}_1 \overline{t} \to t \overline{t} t \overline{t} + 2 \widetilde{\chi}_1^0$. We can further increase the jet count for squark production through $\widetilde{q} \to q \widetilde{g}$ with $\widetilde{q} + \widetilde{g} \to q + t \overline{t} t \overline{t} + 2 \widetilde{\chi}_1^0$ or $\widetilde{q} + \widetilde{q} \to qq + t \overline{t} t \overline{t} + 2 \widetilde{\chi}_1^0$. Therefore, we find the ATLAS multijet SUSY searches~\cite{ATLAS-CONF-2012-033,ATLAS-CONF-2012-037,ATLAS-CONF-2012-103,ATLAS-CONF-2012-109} employing the larger cut on jet $p_T$ to represent an imperfect strategy for \fsu5 discovery.  Given that a very large $p_T$ cut on jets reduces the sensitivity of the SUSY searches to \fsu5 multijets, we fear production of an empty cupboard if a strategy closer to the $\ge$9 jets and $p_T > 20$ GeV suggestion cannot be consistently implemented.  Considering that the chief background competition to SUSY multijets in data with such a low $p_T$ cut on jets is QCD and $t \overline{t}$, the recent addition of a lepton requirement to both the ATLAS~\cite{ATLAS-CONF-2012-073,ATLAS-CONF-2012-140,ATLAS-CONF-2012-166,Aad:2012naa} and CMS~\cite{CMS-PAS-SUS-12-023} arsenal is significant for \fsu5 discovery. The suppression of QCD by the lepton has allowed the $p_T$ cut to be lowered to $p_T >$ 20--30 GeV in multijet searches by both Collaborations, maximizing the sensitivity of the multijet searches to \fsu5. This in effect retains as many hadronic jets as possible, eliminating the unfortunate outcome of discarded SUSY multijets in the context of \fsu5. The intent of our in-depth analysis here is to examine very closely the recent offering of SUSY searches by ATLAS and CMS studying multijet events with the requirement of a lepton, thus permitting a low $p_T >$ 20--30 GeV cut on jets.

Our original implementation of a Monte-Carlo collider-detector simulation utilized the {\tt MadGraph}~\cite{Stelzer:1994ta,MGME} program suite, including the {\tt MadEvent}~\cite{Alwall:2007st}, {\tt PYTHIA}~\cite{Sjostrand:2006za} and {\tt PGS4} chain. Despite the excellent treatment of the entire collider-detector simulation process by the {\tt MadGraph} suite, this platform required an adjustment to the final event counts by a calibration factor~\cite{Li:2012tr} computed from a common benchmark point in order to scale our Monte-Carlo event counts to the Collaboration results. This factor approximately accounted for the exclusion of inclusive up to next-to-leading order (NLO) QCD contributions in the {\tt MadEvent} production cross-section calculations. We now eliminate the necessity of the calibration factor by integrating {\tt PROSPINO~2.1}~\cite{Beenakker:1996ch,Beenakker:1997ut,Beenakker:1996ed} into our Monte-Carlo suite and apply the computed K-factor to each pair-production channel independently, thereby incorporating all correction terms up to NLO in the strong coupling constant. The net effect is an improvement in accuracy in our event counts without the need to resort to application of a single {\it ad hoc} calibration factor to each search to compensate for the partial QCD contributions. Our common benchmark statistics now yield efficiencies after all cuts that agree with the Collaboration results to a baseline precision of around 20\%, a scale comparable to known limitations of our detector simulation and corrections for resummation of soft gluon emission at next-to-leading-logarithmic accuracy (NLO+NLL), which account for possibly an additional 20\% of the total production cross-section for gluino masses around 1 TeV~\cite{Beenakker:2011fu}, depending on the final states.  This small margin of error will have only a minor effect on the SUSY mass scale best fits computed from our multi-axis $\chi^2$ fitting procedure.

Emphasizing again, our original SUSY discovery methodology for \fsu5 advocated a rather large cut on hadronic jets, coupled with a rather small cut on jet $p_T$. By examining the effect of the flippon multiplets on the LHC final states, we derived optimum data cuts for No-Scale \fsu5 of $\ge$9 jets and $p_T > 20$ GeV~\cite{Li:2011hr,Maxin:2011hy}. At the time of our suggested strategy in early 2011, no SUSY searches by either Collaboration had yet surfaced with such a large cut on jets or low cut on jet $p_T$. Later in 2011, ATLAS released the first LHC SUSY search that employed large multijet events through all-hadronic searches, with cuts on jets of $\ge$7, $\ge$8, and $\ge$9~\cite{ATLAS-CONF-2012-037}. This only fulfilled half of our \fsu5 discovery vision though, as jet $p_T$ cuts of $p_T > 55$ GeV and $p_T > 80$ GeV were established at these high levels to suppress the QCD background infiltration. The larger $p_T > 80$ GeV cut permitted likely suppression of \fsu5 in the LHC data as a result of the emission of a soft top quark in $\widetilde{g} \to \widetilde{t}_1 \overline{t}$, due to the potential of the light stop in \fsu5 living near an off-shell to on-shell transition. Such soft jets could get discarded by a large cut on jet $p_T$, masking potential signals emanating from an \fsu5 framework. If in the future there are observed consistent larger event excesses with smaller $p_T$ cuts on jets in leptonic SUSY searches, though concurrently no substantial excess events are observed in the all--hadronic SUSY searches with larger $p_T$ cuts on jets, then this could potentially be regarded as one important piece of evidence that the gluino mass may indeed reside near this on-shell transition in \fsu5, where the jets are expected to be softer.

In 2012, we saw the first ATLAS and CMS SUSY searches implementing a low $p_T$ cut near the $p_T > 20$ GeV level we first suggested, with ATLAS applying a cut at $p_T > 20-25$ GeV~\cite{ATLAS-CONF-2012-073,ATLAS-CONF-2012-140,ATLAS-CONF-2012-166,Aad:2012naa} for both the 7 TeV and 8 TeV data, while CMS has placed a cut at $p_T > 30$ GeV in their 8 TeV light stop search data~\cite{CMS-PAS-SUS-12-023}. The CMS 7 TeV light stop search version inserted the $p_T$ cut at $p_T > 40$ GeV~\cite{:2012pca}, which we consider to be too high for leptonic SUSY searches which are already suppressed by the mere requirement of the lepton, thus we shall not consider that search here. All of these low $p_T$ cuts are enabled by the requirement of a lepton, permitting a large suppression of the QCD background and thus empowering a genuine search for SUSY in events with the large number of soft jets expected in \fsu5. All but two of these low $p_T$ cut SUSY searches target light stop pair-production by both LHC Collaborations utilizing both 7 TeV and 8 TeV data, the exceptions being ATLAS 7 TeV multijet~\cite{ATLAS-CONF-2012-140} and razor~\cite{Aad:2012naa} searches focused on gluino pair-production events. The presence of both ATLAS and CMS SUSY searches with very similar data cutting strategies raises the stakes considerably, particularly in light of static data cuts within the expansion of beam collision energy to 8 TeV from the starting point of 7 TeV, as intrinsic consistency between search results across detector technology and beam collision energy can magnify any observable correlations. This is the scope of our work here, to engage in a first-time comprehensive examination of these soft multijet SUSY searches with a lepton, quite close to our original favored methodology for \fsu5.

The promise of \fsu5 relies not only on the correlations across detectors and beam energies, but also intriguing consistency with other phenomenological experiments external to LHC also mining data for beyond the Standard Model (BSM) physics. We have already demonstrated consistency amongst the early 8 TeV LHC all-hadronic data and other BSM experiments~\cite{Li:2012mr}. We referred to this convergence of best fits of all BSM experiments as Primordial Synthesis~\cite{Li:2012mr}, and as we shall update and show here, the unification amongst all the BSM experiments with the LHC SUSY search data in Primordial Synthesis becomes even further strengthened by inclusion of the leptonic SUSY searches.

\section{The No-Scale \fsu5 Model}

No-Scale \fsu5 (See
Refs.~\cite{Li:2010ws,Li:2010mi,Li:2010uu,Li:2011dw,Li:2011hr,Maxin:2011hy,Li:2011xu,Li:2011in,
Li:2011gh,Li:2011rp,Li:2011fu,Li:2011xg,Li:2011ex,Li:2011av,Li:2011ab,Li:2012hm,Li:2012tr,Li:2012ix,Li:2012yd,Li:2012qv,Li:2012jf,Li:2012mr}
and all references therein) is built upon the tripodal foundation of the Flipped $SU$(5)~\cite{Barr:1981qv,Derendinger:1983aj,Antoniadis:1987dx} grand unified theory (GUT), two pairs of hypothetical TeV-scale vector-like flippon multiplets of mass $M_V$ derived from local F-Theory model building~\cite{Jiang:2006hf,Jiang:2009zza,Jiang:2009za,Li:2010dp,Li:2010rz}, and the dynamically established boundary conditions of No-Scale supergravity~\cite{Cremmer:1983bf,Ellis:1983sf, Ellis:1983ei, Ellis:1984bm, Lahanas:1986uc}. Supersymmetry naturally resolves the gauge hierarchy problem in the Standard Model (SM), and suggests (given $R$ parity conservation) the lightest supersymmetric particle (LSP) as a suitable cold dark matter candidate. However, since we do not observe mass degeneracy of the superpartners, SUSY must be broken around the TeV scale. In GUTs with
gravity mediated supersymmetry breaking, referred to as supergravity models, we can fully characterize the supersymmetry breaking soft terms by four universal parameters: gaugino mass $M_{1/2}$, scalar mass $M_0$, trilinear soft term $A$, and the low energy ratio of Higgs vacuum expectation values (VEVs) $\tan\beta$, plus the sign of the Higgs bilinear mass term $\mu$. In the most simple No-Scale scenario,
$M_0 = A = B_{\mu} = 0$ at the unification boundary, while the entire array of low energy SUSY breaking soft-terms evolve down
from the sole non-zero parameter $M_{1/2}$. As a result, the particle spectrum will be proportional to $M_{1/2}$ at leading order,
rendering the bulk ``internal'' physical properties invariant under an overall rescaling. 

The matching condition between the low-energy value of the $B_\mu$ parameter that is required by electroweak symmetry breaking (EWSB) and the high-energy $B_\mu = 0$ boundary is extremely difficult to reconcile under the RGE running. The current solution relies on modifications to the $\beta$-function coefficients that are generated by the flippon loops. Naturalness in view of the gauge hierarchy and $\mu$ problems do suggest that the flippon mass $M_{\rm V}$ should be of the TeV order. Avoiding a Landau pole for the strong coupling constant restricts the set of vector-like flippon multiplets which may be given a mass in this range to only two constructions with flipped charge assignments, which have been explicitly realized in the $F$-theory model building context~\cite{Jiang:2006hf,Jiang:2009zza, Jiang:2009za}. In either case, the (formerly negative) one-loop $\beta$-function coefficient of the strong coupling $\alpha_3$ becomes precisely zero, flattening the RGE running, and generating a large
gap between the large $\alpha_{32} \simeq \alpha_3(M_{\rm Z}) \simeq 0.11$ and the much smaller $\alpha_{\rm X}$ at the scale $M_{32}$ of the intermediate flipped $SU(5)$ unification of the $SU(3)_{\rm C} \times SU(2)_{\rm L}$ subgroup. This facilitates a very important secondary running phase up to the final $SU(5) \times U(1)_{\rm X}$ unification scale, which may be elevated by 1-2 orders of magnitude
into adjacency with the Planck mass, where the $B_\mu = 0$ boundary condition fits like hand to glove~\cite{Ellis:2001kg,Ellis:2010jb,Li:2010ws}.

\section{No-Scale \fsu5 Phenomenology}\label{sect:pheno}

The identical flippon induced perturbation to the RGE unification framework of \fsu5 that facilitated a consistent application of the No-Scale boundary conditions near the Planck mass also yields a key phenomenological signature. The flat RGE evolution of the $SU(3)_C$ gaugino mass $M_3$, which 
mirrors the flatness of the $\beta$-coefficient $b_3 = 0$, suppresses the standard logarithmic mass enhancement at low-energy
and generates a SUSY spectrum mass ordering of $M(\widetilde{t}_1) < M(\widetilde{g}) < M(\widetilde{q})$,
where the light stop $\widetilde{t}_1$ and gluino $\widetilde{g}$ are both considerably lighter than all other squarks.
This phenomenologically favorable hierarchy produces a distinctive event topology initiated by the pair-production
of heavy first or second generation squarks $\widetilde{q}$ and/or gluinos $\widetilde{g}$ in the hard scattering
process, with the heavy squark most probably yielding a quark-gluino pair $\widetilde{q} \rightarrow q \widetilde{g}$.
The gluino then has only two main channels available in the cascade decay,
$\widetilde{g} \rightarrow \widetilde{t}_1 \overline{t}$ or $\widetilde{g} \rightarrow q \overline{q} \widetilde{\chi}_1^0$,
with $\widetilde{t}_1 \rightarrow t \widetilde{\chi}_1^0$ or $\widetilde{t}_1 \rightarrow b \widetilde{\chi}_1^{\pm}$.
The stop-top channel becomes dominant as $M_{1/2}$ increases, achieving 100\% for $M_{1/2} \ge 729$ GeV with the condition $M(\widetilde{g}) - M(\widetilde{t}_1) \ge M(t)$. Each gluino produces 2--6 hadronic jets, with the maximum of six jets realized in the gluino-mediated stop decay, so that a single gluino-gluino pair-production event can net 4--12 jets. Given fragmentation processes, the final event is
characterized by an unmistakable SUSY signal of high-multiplicity jets.

The region of the parameter space near the off-shell to on-shell transition for gluino-mediated light stops incurs a distinctive effect. As discussed above, the transition at $M_{1/2} = 729$ GeV possesses the attribute $M(\widetilde{g}) - M(\widetilde{t}_1) = M(t)$, thus, in the gluino mediated decay $\widetilde{g} \to \widetilde{t}_1 t$, the top quark emitted directly from the gluino could be very soft at $M_{1/2} \sim 729$ GeV, leading to soft hadronic jets via $t \to b W^{\pm} \to b q \overline{q}$ for multijets. As a result, in our 4-top scenario $\widetilde{g} + \widetilde{g} \to t \overline{t} t \overline{t} + 2 \widetilde{\chi}_1^0$, half of the hadronic jets could be soft, if indeed the SUSY mass scale resides near this off-shell to on-shell transition in No-Scale \fsu5. This prospect imparts a great consequence on where to locate the $p_T$ cut on hadronic jets, so as to preserve as many of these soft jets as possible. Bear in mind that these soft final states are a function of the SUSY mass scale position in reference to the on-shell transition at $M_{1/2} = 729$ GeV. Therefore, we present the case that if the LHC production is consistently beyond expectations with lower $p_T$ cuts of $p_T > 20-30$ GeV for leptonic searches and $p_T > 55$ GeV for all-hadronic searches, though conversely the LHC production is consistent with background expectations with $p_T$ cuts sufficiently larger than those noted above, then this could in fact supply an essential clue indicating that the physically probed SUSY mass scale intrinsically possesses these soft hadronic jets near this on-shell transition of $M_{1/2} = 729$.

Our preferences for the model rests not only on attractive 7 TeV and 8 TeV collider signals, but also on complementary distinctive phenomenology~\cite{Li:2010ws,Li:2010mi,Li:2010uu,Li:2011dw,Li:2011xu,Li:2011in,Li:2011gh,Li:2011xg,Li:2011ex,Li:2011ab,Li:2012yd,Li:2012qv,Li:2012jf,Li:2012mr}. Key are the higher-order corrections from the vector-like flippon multiplets, enhancing the minimal supersymmetric standard model (MSSM) Higgs boson mass by an additional 3--4 GeV~\cite{Li:2011ab}, precisely the magnitude necessary to yield a total Higgs boson mass of 125--126 GeV~\cite{Li:2011ab,Li:2012qv,Li:2012jf}. This feat must be regarded in parallel with the ability to generate a testably light SUSY spectrum at the LHC, a characteristic distinction of No-Scale \fsu5. While the typical MSSM Higgs boson mass of alternative SUSY models, such as minimal supergravity (mSUGRA) and the constrained MSSM (CMSSM), falls noticeably shy of a 125 GeV Higgs boson when attempting to maintain sub 2 TeV gluinos and squarks, on the contrary, by virtue of the flippons, the \fsu5 mass hierarchy $M(\widetilde{t}_1) < M(\widetilde{g}) < M(\widetilde{q})$ of sub 1 TeV light stops and gluinos and sub 2 TeV squarks are within the reach of both the 7 TeV and 8 TeV LHC, and can concurrently produce the now empirically measured 125--126 GeV Higgs boson.

The model's phenomenology also supports global consistency with all currently running BSM experiments, a homogeneity we have characterized as Primordial Synthesis~\cite{Li:2012mr}. The model space satisfying what we refer to as the bare-minimal constraints~\cite{Li:2011xu}, further circumscribed by a 124--127 GeV Higgs boson $m_h$, identifies a narrow strip of model space from which to explore new physics. The resulting strip represents a seldom seen convergence of $124 \le m_h \le 127$ GeV~\cite{:2012gk,:2012gu,Aaltonen:2012qt}, relic density within the 7-year WMAP limits $0.1080 \le \Omega h^2 \le 0.1158$~\cite{Komatsu:2010fb}, and a top quark mass $m_t$ within the world average $172.2 \le m_t \le 174.4$ GeV~\cite{:1900yx}. To forge the Primordial Synthesis, we computed upon this strip of model space the proton lifetime $\tau_p$, the rare decay processes $b \to s \gamma$, ${\rm (g_{\mu}-2)/2}$, and \bs0, the spin-independent cross-section $\left\langle \sigma _{SI}\right\rangle$, and the annihilation cross-sections $\left\langle \sigma v \right\rangle_{\gamma \gamma}$ and $\left\langle \sigma v \right\rangle_{\gamma Z}$~\cite{Li:2012mr}. The uncommon alliance shared by all these experimental observables in No-Scale \fsu5 amplifies our perception that the latent correlations previously uncovered in the multijet data~\cite{Li:2012tr,Li:2012ix}, which are to be further broadened here, may indeed represent a genuine probing of SUSY physics by the 7 and 8 TeV LHC operational phases.  We suggest that the No-Scale \fsu5 correlations uncovered in the LHC collider analyses should not be scrutinized in isolation, but alternatively in the global context of the model's entire array of phenomenological attributes, which in totality presents a rather compelling case for engaging in a deeper understanding of what may lie at the root of the provocative clustering of ATLAS and CMS SUSY searches around a single SUSY mass scale.

We call to special attention a particularly important recent observation in the LHCb~\cite{Aaij:2013aka} and CMS~\cite{Chatrchyan:2013bka} \textit{Br}(\bs0) measurements. The search for this rare decay using 2.0 \fb at 8 TeV and 1.0 \fb at 7 TeV by LHCb and 20 \fb at 8 TeV and 5 \fb at 7 TeV by CMS has uncovered the first evidence of the process \bs0, where a 4.0--4.3$\sigma$ excess above the background expectation was observed. The results indicate a branching fraction of $1.9 \times 10^{-9} \le$ \textit{Br}(\bs0) $\le 4.0 \times 10^{-9}$. We have evaluated this process within the entire No-Scale \fsu5 viable model space~\cite{Li:2012yd}, finding the parameter space constraints $3.4 \times 10^{-9} \le$ Br(\bs0) $\le 4.0 \times 10^{-9}$ for $400 \le M_{1/2} \le 900$ GeV. As a result of the rather small globally permitted range of $19.4 \lesssim {\rm tan}\beta \lesssim 23$~\cite{Li:2011xu}, the \fsu5 SUSY contribution to \textit{Br}(\bs0), which is proportional to the sixth power of tan$\beta$, is much smaller than the effect anticipated within the SM. While the LHCb and CMS observations are consistent with the branching ratio expected within the SM, which should further constrain or invalidate more SUSY models, the case for \fsu5 is actually bolstered by the new measurements.

For a Higgs boson mass of about 125.5 GeV, the Higgs quartic coupling $\lambda$ in the SM will become negative around $10^{10}$ GeV~\cite{EliasMiro:2011aa}, which is the stability bound on the ultraviolet scale. Since our Universe could live in a meta-stable vacuum, the SM can still satisfy the meta-stability bound~\cite{EliasMiro:2011aa}. It is interesting that in the supersymmetric SMs which include our ${\cal F}$-$SU(5)$ model, the stability bound is naturally satisfied due to supersymmetry. In particular, the natural supersymmetric SMs predict the light SM-like Higgs boson mass around 120 GeV, which is lifted to 125.5 GeV in the ${\cal F}$-$SU(5)$ model due to the vector-like flippons~\cite{Huo:2011zt, Li:2011ab}.

In the SM, the anomalous magnetic moment $a_{\mu} \equiv {\rm (g_{\mu}-2)/2}$ of the muon can be divided into the electromagnetic, hadronic, and electroweak contributions. The hadronic contribution gives the largest theoretical uncertainty, where its accurate evaluations must rely on the following experimental measurements:
(1) $e^{+} e^{-} \to {\rm hadrons}$ and $e^{+} e^{-} \to \gamma {\rm hadrons}$;
(2) $\tau^{\pm} \to \nu \pi^{\pm} \pi^0$.
Using the electron data, we have
$\Delta a_{\mu} \equiv a_{\mu}({\rm exp})-a_{\mu}({\rm SM})$ is $28.7 \pm 8.0 \times 10^{-10} $~\cite{Davier:2010nc} and $26.1 \pm 8.0 \times 10^{-10}$~\cite{Hagiwara:2011af}, which correspond respectively to $3.6 \sigma$ and $3.3 \sigma$ discrepancies.
Recently, including the complete tenth-order QED contributions and improving the eighth-order QED contributions, one obtained $\Delta a_{\mu} = 24.9 \pm 8.7 \times 10^{-10}$~\cite{Aoyama:2012wk}, which corresponds to a $2.9 \sigma$ discrepancy.
Using the $\tau$ data, we have
$\Delta a_{\mu} = 19.5 \pm 8.3 \times 10^{-10}$~\cite{Davier:2010nc}, which corresponds to $2.4 \sigma$ discrepancy.
Therefore, considering the average, we will take $\Delta a_{\mu} = 22.5 \pm 8.5 \times 10^{-10}$ at the 1$\sigma$ deviation.
The $2 \sigma$ deviation of $\Delta a_{\mu} = 22.5 \pm 17 \times 10^{-10}$ to this experimental central value provides an upper limit in the No-Scale \fsu5 parameter space of $M_{1/2} \simeq 850$ GeV.

\section{The 7 \& 8 TeV Leptonic SUSY Multijet Searches at the LHC}\label{sect:susy}

We have previously studied SUSY multijets in the context of an all-hadronic requirement for gluino and squark pair-production events in an \fsu5 framework~\cite{Li:2012tr,Li:2012ix,Li:2012mr}. We now turn our examination toward SUSY multijet events requiring a single lepton, thus constituting a completely statistically independent data sample from the all-hadronic observations. We shall outline our procedure in great detail in the next section for quantifying the statistical independence of all leptonic searches, and weighting each accordingly. Three 4.7 \fb 7 TeV ATLAS leptonic SUSY searches~\cite{ATLAS-CONF-2012-073,ATLAS-CONF-2012-140,Aad:2012naa} are investigated here, in addition to an ATLAS 13 \fb 8 TeV leptonic SUSY search~\cite{ATLAS-CONF-2012-166} and CMS 9.7 \fb leptonic SUSY search~\cite{CMS-PAS-SUS-12-023}.

The ATLAS 7 TeV multijet SUSY search of Ref.~\cite{ATLAS-CONF-2012-140} preserves only those events with at least 7 jets and $p_T > 25$ GeV, targeting gluino and squark pair-production, with a single lepton in the final state. On the other hand, the data cutting methodology of the ATLAS 7 TeV multijet SUSY search of Ref.~\cite{ATLAS-CONF-2012-073} is designed so as to capture a larger sample of light stop pair-production events, where the final multijet states must also be coupled with a single electron or muon, along with $\ge 4$ jets, $\ge 1$ b-jet, $p_T > 25$ GeV, and ${\rm E_T^{Miss}} > 275$ GeV in the SRE search region. A close scrutiny of Figure (5) in ATLAS Ref.~\cite{ATLAS-CONF-2012-073} clearly illustrates that all the excesses emanate from events with 9--10 jets, in both the electron and muon channels. Hence, even though the over-production beyond expectations for each of these two 7 TeV leptonic SUSY searches are in events with 7 or more jets, we shall show that these two searches are reasonably statistically independent, with an overlap of events of only about 26\%. Yet, despite this small measure of intersection, the excess production corresponds to the same SUSY mass scale. This minor intersection of events notwithstanding, the ATLAS~\cite{ATLAS-CONF-2012-166} and CMS~\cite{CMS-PAS-SUS-12-023} 8 TeV searches are certainly statistically independent from these 7 TeV searches~\cite{ATLAS-CONF-2012-073,ATLAS-CONF-2012-140}, as well as independent from each other.

The ATLAS 7 TeV multijet+lepton razor search~\cite{Aad:2012naa} is an important contribution to the SUSY search landscape. In the razor methodology, two mega-jets are constructed from all final-state particles using a combinatorial scheme. The vector sum of the mega-jet momenta can then identify when ${\rm E_T^{Miss}}$ originates from a detector defect or mismeasurement (small vector sum), as opposed to whether the ${\rm E_T^{Miss}}$ is genuine (large vector sum), providing good discrimination between SUSY events and backgrounds without real ${\rm E_T^{Miss}}$. This ATLAS version of razor is intended to acquire gluino pair-production events with gluino-mediated light stops to $t \overline{t}$. In our work here, we examine the single lepton search regions, which are segregated into b-tag and b-veto, for both an electron and muon in the final state. The data cuts are certainly keeping with the spirit of our goal in this paper, with a hadronic jet count of $\ge$5 jets and hadronic jet cut of $p_T > 20$ GeV, justifying its presence as an essential ingredient in our analysis here. The tight cuts are implemented as well, with $M'_R > 600$ GeV for the b-veto and $M'_R > 1000$ GeV for the b-tag. The 7 TeV razor b--tag searches are essentially statistically independent from the 7 TeV SRE and multijet+lepton searches, with an overlap of events of only 15\% with SRE and 36\% with multijets+lepton. Likewise, the razor b--veto searches have only a very small intersection of 1\% with SRE and 6\% with multijets+lepton. As we shall show in our Monte-Carlo results, the excesses in the ATLAS data for the b-tag are likewise matched by events in our \fsu5 simulation, while the lack of any excesses in the ATLAS data for the b-veto are equally complemented by only a fraction of an event in the \fsu5 simulation. Stated otherwise, when events are needed, \fsu5 provides events, though when no events are needed, \fsu5 does not deliver events.

The ATLAS~\cite{ATLAS-CONF-2012-166} 8 TeV leptonic multijet results focus on light stop pair-production, in concert with the ATLAS 7 TeV light stop search of Ref.~\cite{ATLAS-CONF-2012-073}. The ATLAS leptonic 13 \fb 8 TeV SUSY search duplicates the 7 TeV search region SRE ($\ge 4$ jets, $\ge 1$ b-jet, $p_T > 25$ GeV, ${\rm E_T^{Miss}} > 275$ GeV), permitting a direct comparison of SUSY mass scale best fits across beam collision energies. The SRD search region wholly contains SRE, though when subtracting out the SRE space from SRD, the remaining portion is below the \fsu5 region of interest, so we shall not consider SRD here. Three additional search regions are studied for different LSP scenarios: (i) SRtN1 for models where $M(\widetilde{t}_1) \gtrsim M(t) + M(\widetilde{\chi}_1^0)$; (ii) SRtN2 for models with large $M(\widetilde{\chi}_1^0)$; and (iii) SRtN3 for models with very large $M(\widetilde{t}_1)$. The phenomenologically favored region of the No-Scale \fsu5 parameter space has $130 \gtrsim M(\widetilde{\chi}_1^0) \gtrsim 180$ GeV and $700 \gtrsim M(\widetilde{t}_1) \gtrsim 950$ GeV, thus, of all the ATLAS searches in Ref.~\cite{ATLAS-CONF-2012-166}, \fsu5 would be most sensitive to the SRE and SRtN3 search regions. Indeed, the ATLAS observations show that SRE and SRtN3 have the largest signal significances, with the remaining regions either in a deficit or only a small excess.  We discover a non-trivial overlap of events amongst these searches, with SRE statistically dependent with SRtN1, SRtN2, and SRtN3 at levels of 55\%, 70\%, and 62\%, respectively. Also, the SRtN2 and SRtN3 regions intersect at 80\%. The region SRtN1 is essentially independent with SRtN2 and SRtN3 at 35\% and 25\%, respectively. However, each of these searches provides a different perspective on the probed region of the parameter space, so we shall retain all four searches for the multi-axis $\chi^2$ fitting procedure.

The CMS 9.7 \fb 8 TeV leptonic SUSY search~\cite{CMS-PAS-SUS-12-023} also applies data cuts of $\ge 4$ jets and $\ge 1$ b-jet, while slightly elevating the $p_T$ cut on jets to $p_T > 30$ GeV. CMS, however, segregates the search regions into six regions, with increasing minimum values of ${\rm E_T^{Miss}}$, where here we shall focus on the CMS search regions SRD (${\rm E_T^{Miss}} > 250$ GeV), SRE (${\rm E_T^{Miss}} > 300$ GeV), and SRF (${\rm E_T^{Miss}} > 350$ GeV). This allows a discrete binning of ${\rm E_T^{Miss}}$ to more carefully isolate specific characteristics of the missing energy profile. It further permits the large statistical dependence of these three regions to be nullified since SRE is a wholly contained subset of SRD, and likewise with SRF a subset of SRE. The close connection of the 8 TeV ATLAS SRE and CMS light stop data cutting strategies offers a rare opportunity to directly compare SUSY mass scale best fits in the context of No-Scale \fsu5 across detector technologies.

\section{Statistical Independence of SUSY Search Strategies}

A substantial variety of novel kinematic and particle identification discriminants
have been devised and implemented by the LHC collaborations for the purpose of
emphasizing supersymmetric processes in preference to the Standard Model background.
By alternately filtering a common catalog of Monte Carlo collider-detector event
simulations against a bank of discrete selection cut strategies, it is possible to
systematically quantify the pairwise overlap between any two basis elements of the
compound search space.  The degree of cross-correlation may be represented
for a set of indexed searches by a symmetric matrix,
\begin{equation}
C_{i,j} \equiv \frac{N_{i \cap j}}{\sqrt{N_i N_j}}\, ,
\end{equation}
where $N_i$ and $N_j$ are the count of surviving events in the $i^{\rm th}$
and $j^{\rm th}$ channels, respectively, and $N_{i \cap j}$ is the numerical
intersection of events accepted by both strategies.
In our current treatment, this is complicated slightly by the fact that
small groups of related production channels (such as squark-squark, stop-antistop,
squark-gluino, gluino-gluino, etc.) are generated at the Monte Carlo level
and passed through the selection cuts in isolation, such that they may
each be rescaled by the corresponding
{\tt PROSPINO~2.1}~\cite{Beenakker:1996ch,Beenakker:1997ut,Beenakker:1996ed}
NLO K-factor.  In the circumstance that individual production channels $k$
must be recombined from split correlation matrix elements $C_{i,j}^k$,
the cumulative correlation factor for the $i^{\rm th}$ and $j^{\rm th}$
channels is represented as a sum
\begin{equation}
C_{i,j} \equiv \sum_k C_{i,j}^k \sqrt{\left( P_i^k P_j^k \right)}\, ,
\end{equation}
where $P_i^k$ and $P_j^k$ are the relative fractions of the total events
passing selection $i$ or $j$ that are attributable to the production mode $k$.

As an example scenario, the following matrix represents the correlation
for \fsu5 events satisfying the selection cut criteria of those four 8 TeV
13 \fb ATLAS lepton and jet stop squark SUSY search channels~\cite{ATLAS-CONF-2012-166}
from which the second pane of Figure (\ref{fig:chi_square}) is composed. These correlations apply specifically at the benchmark $M_{1/2} = 800$~GeV, which is adjacent to the best fit for this set of searches.
\begin{equation}
C_{\rm ATLAS} =
\begin{pmatrix}
1.0 & 0.555 & 0.701 & 0.622 \\
0.555 & 1.0 & 0.353 & 0.255 \\
0.701 & 0.353 & 1.0 & 0.796 \\
0.622 & 0.255 & 0.796 & 1.0
\end{pmatrix} 
\label{eq:catlas}
\end{equation} 
The adopted channel ordering for indices ${\rm i}\ldots {\rm iv}$ is
i) SRE, ii) SRtN1, iii) SRtN2, and iv) SRtN3. A description of these explicit signal regions is discussed in Section~\ref{sect:susy}.
It should be observed that the considered searches are relatively strongly
cross coupled, providing a rather ideal candidate for analysis and treatment.

A mechanism, such as that described, for concretely gauging the level of statistical
interdependence of various search channels is essential for prudently interpreting the
strength of conclusions based upon a combination of those channels.  Even more
beneficial, however, would be a mechanism for statistically weighting individual 
channels that bear some component of overlap, such that degeneracies were projected out,
retaining an independent subspace basis spanning a reduced count of effective degrees of
freedom.  A candidate procedure for the realization of that goal is presented in the
present section, which will be referenced to justify combination
of the described ATLAS searches in a subsequent $\chi^2$ analysis.

By way of motivation, we shall first consider the two extreme boundary
cases confronting an $N$-dimensional search space.  In the former case,
we will assume a perfect independence (I) of all channels from the outset,
such that the correlation is represented by a unit diagonal matrix $C_{\rm I}$.
\begin{equation}
C_{\rm I}
\equiv 
\begin{pmatrix}
1 & 0 & \cdots & 0 \\\vspace{-5pt}
0 & 1 & \cdots & 0 \\
\vdots & \vdots & \ddots & 0 \\
0 & 0 & 0 & 1
\end{pmatrix} 
\end{equation}
In the latter case,
we will assume a perfect degeneracy (D) of all channels,
such that all elements of the the correlation matrix $C_{\rm D}$
are equal to one.
\begin{equation}
C_{\rm D}
\equiv
\begin{pmatrix}
1 & 1 & \cdots & 1 \\\vspace{-5pt}
1 & 1 & \cdots & 1 \\
\vdots & \vdots & \ddots & 1 \\
1 & 1 & 1 & 1
\end{pmatrix} 
\end{equation} 
A straightforward measure of the overall search entanglement
is provided by the magnitude-square of the vector ${\left|W \right\rangle}_j$
representing the search weighting coefficients, employing 
the correlation matrix $C_{i,j}$ as a metric tensor for the inner product.
Nominally, each search is weighted at the unit level, {\it i.e.}
\begin{equation}
\left|W \right\rangle \Rightarrow \left| \mathbb{1} \right\rangle
\equiv
\begin{pmatrix}
1 \\\vspace{-5pt}
1 \\
\vdots \\
1 \\
\end{pmatrix}\, , 
\end{equation} 
and the corresponding scalar contractions of the two example cases reduce simply to
\begin{equation}
\left\langle \mathbb{1} \right| C_{\rm I} \left| \mathbb{1} \right\rangle = N
\quad ; \quad
\left\langle \mathbb{1} \right| C_{\rm D} \left| \mathbb{1} \right\rangle = N^2
\, ,
\end{equation}
where $N$ is the dimensionality of the search space.  Of course, this
result is not surprising; it is simply the statement that $N$ orthogonally
directed unit vectors have a net Pythagorean length of $\sqrt{N}$, whereas
$N$ collinear displacements compound as a simple sum.  The magnitude-square
in the former example is a match for the count of $N$ independent degrees
of freedom represented in that case.  For the fully degenerate
example, it is clear that a $1/N$ reduction in the search weight vector
to $\left|W \right\rangle \Rightarrow \left| \mathbb{1} \right\rangle / N$
would appropriate compensate for an $N$-fold over counting.  Moreover, this
modification precisely counters the $N^2$ factor previously observed
in the associated scalar magnitude-square, such that the new result of $1$ 
again reproduces the single independent search tally.  Based upon these observations,
we may speculate more broadly that the effective degree of freedom count $N^{\rm eff}$
associated with a group of $N$ generically correlated searches may be represented
as the scalar contraction
\begin{equation}
N^{\rm eff} \equiv \left\langle W \right| C \left| W \right\rangle\, ,
\label{eq:neffa}
\end{equation}
where the weight vector $\left| W \right\rangle$ is itself expressible
as a linear matrix transformation
\begin{equation}
\left| W \right\rangle \equiv W \left| \mathbb{1} \right\rangle
\label{eq:weights}
\end{equation}
of the unity vector $\left| \mathbb{1} \right\rangle$.
Additionally, it is expected that the sum of assigned weights in the vector
$\left| W \right\rangle$ should likewise correspond to the
reduced effective degree of freedom count, as expressed
by the following scalar product.
\begin{equation}
N^{\rm eff} \equiv \left\langle \mathbb{1} | W \right\rangle =
\left\langle W | \mathbb{1} \right\rangle =
\left\langle \mathbb{1} \right| W \left| \mathbb{1} \right\rangle
\label{eq:neffb}
\end{equation}
Comparing Eqs.~\ref{eq:neffa} and \ref{eq:neffb}, one concludes that
\begin{equation}
C \left| W \right\rangle =
C W \left| \mathbb{1} \right\rangle =
\left| \mathbb{1} \right\rangle\, ,
\end{equation}
or, equivalently, that
\begin{equation}
W = C^{-1}\, .
\end{equation}
The merit of this construction may be tested by application to
several relevant examples.  We begin with the following
parameterization of a 2-dimensional correlation matrix $C_{\varepsilon}$,
which spans the boundary cases of perfect independence ($\varepsilon = 0$)
and complete degeneracy ($\varepsilon = 1$).
\begin{equation}
C_{\varepsilon} =
\begin{pmatrix}
1 & \varepsilon \\
\varepsilon & 1 
\end{pmatrix} 
\end{equation} 
The two eigenvalues of this matrix are real, as guaranteed in general
by Hermiticity.
\begin{equation}
\lambda_\pm = 1 \pm \varepsilon
\quad ; \quad
\left| \lambda_{\pm} \right\rangle =
\begin{pmatrix}
1 \\
\pm 1
\end{pmatrix} / \sqrt{2} 
\end{equation}
The inversion of this matrix is
\begin{equation}
W_{\varepsilon} =
\begin{pmatrix}
1 & -\varepsilon \\
-\varepsilon & 1 
\end{pmatrix} /\, (1-\varepsilon^2)\, ,
\end{equation} 
yielding the following weight vector and effective degree of freedom count.
\begin{equation} 
\left| W_\varepsilon \right\rangle =
\begin{pmatrix}
1 \\
1
\end{pmatrix} /\, (1+\varepsilon) 
\quad ; \quad 
N^{\rm eff}_\varepsilon = \frac{2}{1+\varepsilon}
\end{equation} 
These results do reduce to those expected in the described limits,
although it should be noted that the inversion matrix becomes undefined
for $\varepsilon \Rightarrow 1$, as mandated by the presence of a zero eigenvalue.
Nevertheless, the tuning of $\varepsilon$ from $0$ to $1$ does correspondingly
map the relevant search weighting coefficients from $1$ to $1/2$, and the
effective degrees of freedom from $2$ to $1$.  This behavior persists in the
$N$-dimensional examples; the fully degenerate limit possesses $N-1$ zero-modes
that effectively truncate the cardinality of the search space in that case,
resulting in the equal weighting of all input searches by $1/N$.
We conclude discussion of the weighting method by analyzing an additional
pair of 3-dimensional correlation examples, to be labeled as $C_{\rm A}$ and $C_{\rm B}$. 
\begin{equation}
C_{\rm A} =
\begin{pmatrix}
1 & \nicefrac{1}{2} & \nicefrac{1}{4} \\
\nicefrac{1}{2} & 1 & \nicefrac{1}{2} \\
\nicefrac{1}{4} & \nicefrac{1}{2} & 1
\end{pmatrix}
\quad ; \quad
C_{\rm B} =
\begin{pmatrix}
1 & \nicefrac{3}{4} & \nicefrac{1}{4} \\
\nicefrac{3}{4} & 1 & \nicefrac{1}{2} \\
\nicefrac{1}{4} & \nicefrac{1}{2} & 1
\end{pmatrix}
\label{eq:cab}
\end{equation}
In example {\rm A}, searches i and iii are each relatively strongly correlated
with search ii ($C_{\rm i,ii}=C_{\rm ii,iii}=\nicefrac{1}{2}$),
although in distinct ways, since they are much more weakly
correlated with each other ($C_{\rm i,iii}=\nicefrac{1}{4}$).
Example {\rm B} represents a variation where
the correlation between searches i and ii
is increased by 50\% to $C_{\rm i,ii}=\nicefrac{3}{4}$.
The corresponding weight matrices given by inversion
of the Eq.~(\ref{eq:cab}) correlations, are expressed as follows.
\begin{equation}
\hspace{-2pt}
W_{\rm A} =
\begin{pmatrix}
\nicefrac{4}{3} & \nicefrac{-2}{3} & 0 \\
\nicefrac{-2}{3} & \nicefrac{5}{3} & \nicefrac{-2}{3} \\
0 & \nicefrac{-2}{3} & \nicefrac{4}{3}
\end{pmatrix}
\, ; \,
W_{\rm B} =
\begin{pmatrix}
\nicefrac{12}{5} & -2 & \nicefrac{2}{5}  \\
-2 & 3 & -1 \\
\nicefrac{2}{5} & -1 & \nicefrac{7}{5}
\end{pmatrix}
\end{equation}
From this, the weight vectors and effective degrees of
freedom emerge directly.
\begin{equation}
\left| W_{\rm A} \right\rangle =
\begin{pmatrix}
\nicefrac{2}{3} \\
\nicefrac{1}{3} \\
\nicefrac{2}{3}
\end{pmatrix}
\quad ; \quad 
N^{\rm eff}_{\rm A} = \frac{5}{3} 
\end{equation}
The result for example {\rm A} is intuitive, devaluing
the largely redundant search ii with respect to the symmetric
treatment afforded searches i and iii.
\begin{equation}
\left| W_{\rm B} \right\rangle =
\begin{pmatrix}
\nicefrac{4}{5} \\
0 \\
\nicefrac{4}{5}
\end{pmatrix}
\quad ; \quad 
N^{\rm eff}_{\rm B} = \frac{8}{5} 
\end{equation}
In variation {\rm B}, the slight increase in the search i to ii correlation
creates an interesting tension as the identity 
$\nicefrac{1}{4} + \nicefrac{1}{2} \equiv \nicefrac{3}{4}$ presses against a limit
that is in some sense analogous to the conventional triangle inequality of plane
geometry.  Specifically, if channel ii is 75\% duplicated by i and 50\% duplicated by iii,
then the minimal possible overlap between channels i and iii is given as the difference
of 25\%, which may furthermore only be realized when the union of sets ${\rm i}\cup{\rm iii}$ contains 
the entirety of channel ii's events, rendering it redundant.  Following this truncation,
the weighting of the remaining two channels must be symmetric, and based only on their
mutual pairwise correlation, as is observed.  Despite this apparently substantive
divergence in outcome between cases {\rm A} and {\rm B},
the final count of effective degrees of freedom is virtually identical,
in keeping with the smallness of the perturbation originally enacted within the
correlation matrices.

With a concrete method for the statistical disentanglement of correlated searches
in place, we return to the relevant example at hand, namely treatment of the correlation
matrix in Eq.~(\ref{eq:catlas}) for leptonic ATLAS 8 TeV SUSY searches. The numerical inversion,
which corresponds to the desired matrix of weights, is as follows.
\begin{equation}
W_{\rm ATLAS} =
\begin{pmatrix}
2.591 & -0.926 & -1.077 & -0.518 \\
-0.926 & 1.475 & -0.084 & 0.266 \\
-1.077 & -0.084 & 3.369 & -1.990 \\
-0.518 & 0.266 & -1.990 & 2.839
\end{pmatrix} 
\end{equation}
The associated weight vector and effective degrees of freedom count (reduced from an
actual total of 4) are computed directly.
\begin{equation} 
\left| W_{\rm ATLAS} \right\rangle =
\begin{pmatrix}
0.070 \\
0.732 \\
0.218 \\
0.596
\end{pmatrix}
\quad ; \quad
N^{\rm eff}_{\rm ATLAS} = 1.62
\end{equation}
Note that the very slightly different value for $N^{\rm eff}$ quoted in Figure (\ref{fig:chi_square}) is
based upon an extrapolation of the degrees of freedom tabulated across all $M_{1/2}$
benchmarks, as evaluated at the overall best fit mass for the set of searches.

\section{A SUSY Search Strategy Using The Peak In ${\rm E_T^{Miss}}$}

We have observed in our Monte-Carlo simulations of the No-Scale \fsu5 model space a strongly peaked Poisson-like distribution of the histogram on missing transverse energy ${\rm E_T^{Miss}}$~\cite{Li:2011gh}. Remarkably, the central value appears to robustly correlate with a small multiple of the LSP mass. A full investigation of the theoretical origins and broader generality of the result and persistence of this result across the viable parameter space showed that this multiple is consistently about two in No-Scale \fsu5~\cite{Li:2011gh}. Given the permitted kinetic excess over the jet invariant mass, the trigonometric reduction from extraction of the beam-transverse component, and the potential partial cancellation between the dual neutralino signal with randomly oriented directionality as a result of the multi level cascade decay, it does not seem obvious that this should be the case. We projected that this phenomena could possibly be an effective tool for extraction of the LSP mass from collider level events if the background could be duly suppressed enough to allow extraction of the peak of an event distribution of new physics~\cite{Li:2011gh}.

\begin{table*}[htp]
	\centering
	\caption{Composition of the final states (in \%) after all data cuts for $M_{1/2} = 750$ GeV, in terms of the pair-production channels $\widetilde{t}_1 \widetilde{t}_1,~\widetilde{g} \widetilde{g},~\widetilde{g} \widetilde{q},~\widetilde{q} \widetilde{q}$. Included are all five statistically independent ATLAS and CMS  SUSY search strategies examined and described in this work. Results are based upon our Monte-Carlo collider detector simulation of No-Scale \fsu5 inclusive of all contributions up to next-to-leading order. The SUSY spectrum (in GeV) used in these calculations is $\widetilde{\chi}_1^0 = 154$, $\widetilde{\chi}_1^{\pm} = 329$, $\widetilde{t}_1 = 832$, $\widetilde{g} = 1012$, $\widetilde{b}_1 = 1289$, $\widetilde{t}_2 = 1321$, $\widetilde{b}_2 = 1415$, $\widetilde{u}_R = 1436$, $\widetilde{d}_R = 1488$, $\widetilde{u}_L = 1562$, and $\widetilde{d}_L = 1564$.}
		\begin{tabular}{|c|c||c|c|c|c|c|} \hline
$\sqrt{s}$&$\rm SUSY~ Search$&$\widetilde{t}_1 \widetilde{t}_1$&$\widetilde{g} \widetilde{g}$&$\widetilde{g} \widetilde{q}$&$\widetilde{q} \widetilde{q}$&$\rm other$ \\ \hline \hline	
$\rm ~~~ATLAS~7 ~TeV~~~$&$\rm	~~~Multijet + e^{\pm}~~~	$&$	~~~0.4\%~~~	$&$	~~~45.1\%~~~	$&$	~~~45.6\%	~~~$&$	~~~7.9\%~~~	$&$~~~	1.0\%~~~	$ \\
$\rm ~~~ATLAS~7 ~TeV~~~$&$\rm	~~~Multijet+\mu^{\pm}~~~	$&$	~~~0.5\%~~~	$&$	~~~46.1\%~~~	$&$	~~~44.2\%	~~~$&$	~~~8.1\%~~~	$&$~~~	1.1\%~~~	$ \\
$\rm ~~~ATLAS~7 ~TeV~~~$&$\rm	~~~SRE~ e^{\pm} + \mu^{\pm}~~~	$&$	~~~4.2\%~~~	$&$	~~~52.5\%~~~	$&$	~~~34.2\%	~~~$&$	~~~4.8\%~~~	$&$~~~	4.3\%~~~	$ \\
$\rm ~~~ATLAS~7 ~TeV~~~$&$\rm ~~~Razor~b-tag~ e^{\pm}~~~	$&$	~~~0.8\%~~~	$&$	~~~24.7\%~~~	$&$	~~~59.9\%	~~~$&$	~~~12.9\%~~~	$&$~~~	1.7\%~~~	$ \\
$\rm ~~~ATLAS~7 ~TeV~~~$&$\rm	~~~Razor~b-tag~ \mu^{\pm}~~~	$&$	~~~0.8\%~~~	$&$	~~~25.0\%~~~	$&$	~~~58.8\%	~~~$&$	~~~13.6\%~~~	$&$~~~	1.8\%~~~	$ \\
$\rm ~~~ATLAS~7 ~TeV~~~$&$\rm	~~~Razor~b-veto~ e^{\pm}~~~	$&$	~~~4.2\%~~~	$&$	~~~43.3\%~~~	$&$	~~~39.1\%	~~~$&$	~~~6.5\%~~~	$&$~~~	6.9\%~~~	$ \\
$\rm ~~~ATLAS~7 ~TeV~~~$&$\rm	~~~Razor~b-veto~ \mu^{\pm}~~~	$&$	~~~4.2\%~~~	$&$	~~~42.2\%~~~	$&$	~~~39.6\%	~~~$&$	~~~7.5\%~~~	$&$~~~	6.5\%~~~	$ \\ \hline \hline
$\rm ~~~ATLAS~8 ~TeV~~~$&$\rm	~~~SRE~ e^{\pm} + \mu^{\pm}~~~	$&$	~~~3.6\%~~~	$&$	~~~48.9\%~~~	$&$	~~~37.9\%	~~~$&$	~~~5.6\%~~~	$&$~~~	4.0\%~~~	$ \\
$\rm ~~~ATLAS~8 ~TeV~~~$&$\rm	~~~SRtN1~ e^{\pm} + \mu^{\pm}~~~	$&$	~~~2.7\%~~~	$&$	~~~52.7\%~~~	$&$	~~~35.3\%	~~~$&$	~~~5.1\%~~~	$&$~~~	4.2\%~~~	$ \\
$\rm ~~~ATLAS~8 ~TeV~~~$&$\rm	~~~SRtN2~ e^{\pm} + \mu^{\pm}~~~	$&$	~~~5.8\%~~~	$&$	~~~49.0\%~~~	$&$	~~~34.8\%	~~~$&$	~~~5.1\%~~~	$&$~~~	5.3\%~~~	$ \\
$\rm ~~~ATLAS~8 ~TeV~~~$&$\rm	~~~SRtN3~ e^{\pm} + \mu^{\pm}~~~	$&$	~~~6.4\%~~~	$&$	~~~47.4\%~~~	$&$	~~~35.7\%	~~~$&$	~~~5.4\%~~~	$&$~~~	5.1\%~~~	$ \\ \hline \hline
$\rm ~~~CMS~8 ~TeV~~~$&$\rm	~~~SRD~ e^{\pm}~~~	$&$	~~~2.5\%~~~	$&$	~~~43.3\%~~~	$&$	~~~42.7\%	~~~$&$	~~~8.4\%~~~	$&$~~~	3.1\%~~~	$ \\
$\rm ~~~CMS~8 ~TeV~~~$&$\rm	~~~SRE~ e^{\pm}~~~	$&$	~~~2.4\%~~~	$&$	~~~42.5\%~~~	$&$	~~~43.3\%	~~~$&$	~~~8.8\%~~~	$&$~~~	3.0\%~~~	$ \\
$\rm ~~~CMS~8 ~TeV~~~$&$\rm	~~~SRF~ e^{\pm}~~~	$&$	~~~2.2\%~~~	$&$	~~~40.9\%~~~	$&$	~~~44.7\%	~~~$&$	~~~9.3\%~~~	$&$~~~	2.9\%~~~	$ \\	\hline \hline
$\rm ~~~CMS~8 ~TeV~~~$&$\rm	~~~SRD~ \mu^{\pm}~exclusive~~~	$&$	~~~3.7\%~~~	$&$	~~~49.3\%~~~	$&$	~~~36.2\%	~~~$&$	~~~6.5\%~~~	$&$~~~	4.3\%~~~	$ \\
$\rm ~~~CMS~8 ~TeV~~~$&$\rm	~~~SRE~ e^{\pm}~exclusive~~~	$&$	~~~3.5\%~~~	$&$	~~~51.6\%~~~	$&$	~~~35.0\%	~~~$&$	~~~5.8\%~~~	$&$~~~	4.1\%~~~	$ \\
$\rm ~~~CMS~8 ~TeV~~~$&$\rm	~~~SRF~ e^{\pm}~exclusive~~~	$&$	~~~3.1\%~~~	$&$	~~~48.3\%~~~	$&$	~~~38.4\%	~~~$&$	~~~6.6\%~~~	$&$~~~	3.6\%~~~	$ \\	\hline \hline
$\rm ~~~CMS~8 ~TeV~~~$&$\rm	~~~SRD~e^{\pm}+ \mu^{\pm}~exclusive~~~	$&$	~~~3.2\%~~~	$&$	~~~49.2\%~~~	$&$	~~~37.3\%	~~~$&$	~~~6.2\%~~~	$&$~~~	4.1\%~~~	$ \\
$\rm ~~~CMS~8 ~TeV~~~$&$\rm	~~~SRE~e^{\pm}+ \mu^{\pm}~exclusive~~~	$&$	~~~3.5\%~~~	$&$	~~~48.7\%~~~	$&$	~~~37.5\%	~~~$&$	~~~6.0\%~~~	$&$~~~	4.3\%~~~	$ \\
$\rm ~~~CMS~8 ~TeV~~~$&$\rm	~~~SRF~e^{\pm}+ \mu^{\pm}~exclusive~~~	$&$	~~~3.3\%~~~	$&$	~~~48.2\%~~~	$&$	~~~38.2\%	~~~$&$	~~~6.6\%~~~	$&$~~~	3.7\%~~~	$ \\	\hline

		\end{tabular}
		\label{tab:percentages}
\end{table*}

The No-Scale \fsu5 SUSY spectrum possesses the rather distinct characteristic of leading order $en~ masse$ proportionality to only the single dimensionful parameter $M_{1/2}$. In essence, the internal physics of \fsu5 are largely invariant under a numerical rescaling of only
$M_{1/2}$. As a result, each sparticle mass within the SUSY spectrum can be rescaled by an identical trivial rescaling of only $M_{1/2}$, though the linear relationship between $M_{1/2}$ and each sparticle can vary. Practically speaking, this quite useful property of No-Scale \fsu5 allows for the SUSY spectrum to be approximately determined from only a given value of $M_{1/2}$, or alternatively, from a given value of any particular sparticle mass. Consequently, via determination of the peak in the Poisson-like event distribution for the collider observable ${\rm E_T^{Miss}}$, we can quickly discern that the LSP mass in \fsu5 is about half the peak value of this ${\rm E_T^{Miss}}$ distribution, and then use the rescaling property in reverse to uncover the associated $M_{1/2}$. Once $M_{1/2}$ is given, then the entire SUSY spectrum can be determined. This rescaling of the mass scale $M_{1/2}$ shares an analogous historical bond with the fixing of the Bohr atomic radius $a_0 = 1/(m_e \alpha)$ in terms of the physical electron mass and charge, through minimization of the electron energy~\cite{Feynman}. In each of these two instances, the spectrum scales according to variation in the selected constants, while leaving the relative internal structure of the model intact.

Armed with this useful tool, we could in principle apply this to LHC leptonic and all--hadronic SUSY searches binned in terms of ${\rm E_T^{Miss}}$, in an attempt to uncover any clear peak that may exist in the excess production beyond background expectations, assumed to be new physics in this extraction procedure. We stress that this derivation of the SUSY mass scale is completely orthogonal to the Monte-Carlo method to be illustrated later in this work. A valuable result acquired via measurement of the peak in the Poisson-like distribution of the collider observable ${\rm E_T^{Miss}}$ justifies an expansion in the scope of search strategy options. We thus reserve application of this option in the future in the occurrence of any large excess deviation of events above SM expectations that may occur in a reasonably constrained region of the SUSY mass scale binned in terms of ${\rm E_T^{Miss}}$, which could possibly represent this peak in ${\rm E_T^{Miss}}$ within No-Scale \fsu5.

\section{A Multi-Axis \x2}

The assembly of our multi-axis $\chi^2$ analysis is facilitated by execution on twenty-one benchmark samples an in-depth Monte Carlo collider-detector simulation of all 2-body SUSY processes based on the {\tt MadGraph}~\cite{Stelzer:1994ta,MGME} program suite, including the {\tt MadEvent}~\cite{Alwall:2007st}, {\tt PYTHIA}~\cite{Sjostrand:2006za} and {\tt PGS4} chain, with the {\tt PROSPINO~2.1}~\cite{Beenakker:1996ch,Beenakker:1997ut,Beenakker:1996ed} front-end. The SUSY particle masses are calculated with {\tt MicrOMEGAs 2.4}~\cite{Belanger:2010gh}, applying a proprietary modification of the {\tt SuSpect 2.34}~\cite{Djouadi:2002ze} codebase to run the flippon-enhanced RGEs. We implement a modified version of the default ATLAS and CMS detector specification cards provided with {\tt PGS4} that calls an anti-kt jet clustering algorithm, indicating an angular scale parameter of $\Delta R = 0.4$ (ATLAS) and $\Delta R = 0.5$ (CMS).  The resultant event files are filtered according to a precise replication of the selection cuts specified by the ATLAS and CMS Collaborations in Refs.~\cite{ATLAS-CONF-2012-073,ATLAS-CONF-2012-140,Aad:2012naa,ATLAS-CONF-2012-166,CMS-PAS-SUS-12-023}, employing the script {\tt CutLHCO 2.4}~\cite{Walker:2012vf} to implement the post-processing cuts. The {\tt PGS4} code is modified to implement the 75\% b-tagging efficiency required for implementation of the data cutting strategy employed in Refs.~\cite{ATLAS-CONF-2012-073,Aad:2012naa,ATLAS-CONF-2012-166,CMS-PAS-SUS-12-023}.

\begin{figure*}[htp]
        \centering
        \includegraphics[width=0.85\textwidth]{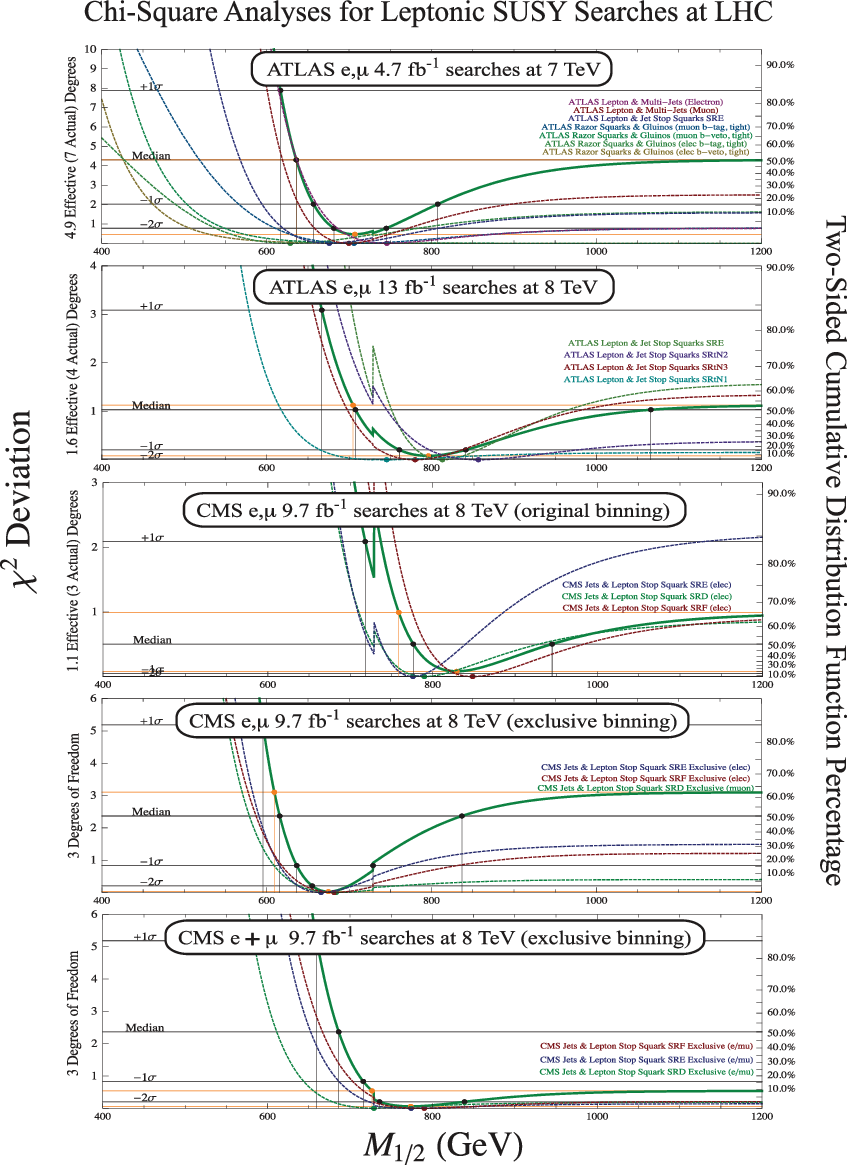}
        \caption{We depict the $\chi^2$ analyses of the 7 TeV ATLAS 4.7 \fb leptonic SUSY searches of Refs.~\cite{ATLAS-CONF-2012-073,ATLAS-CONF-2012-140,Aad:2012naa} (top pane), the 8 TeV ATLAS 13 \fb leptonic SUSY search of Ref.~\cite{ATLAS-CONF-2012-166} (second pane from top), and the 8 TeV CMS 9.7 \fb leptonic SUSY search of Ref.~\cite{CMS-PAS-SUS-12-023} (bottom three panes). The thin dotted lines correspond to the individual $\chi^2$ curves for each event selection for only the case of the nominal number of \fsu5 events, which are summed into the thick green cumulative multi-axis $\chi^2$ curves. The discontinuity shown represents the transition from off-shell to on-shell gluino-mediated light stop production at $M_{1/2} = 729$ GeV. A description of the explicit signal regions depicted here is contained in Section~\ref{sect:susy}.}
        \label{fig:chi_square}
\end{figure*}

Our $\chi^2$ analysis of the 7 TeV and 8 TeV ATLAS and CMS leptonic multijet searches~\cite{ATLAS-CONF-2012-073,ATLAS-CONF-2012-140,Aad:2012naa,ATLAS-CONF-2012-166,CMS-PAS-SUS-12-023}
is depicted in the five panels of Figure (\ref{fig:chi_square}). A description of the explicit signal regions depicted in Figure (\ref{fig:chi_square}) can be found in Section~\ref{sect:susy}. The Cumulative Distribution Function (CDF) percentage labeled on the right-hand axis of each plot is a statistical tool that establishes the fraction of trials (for a fixed number of statistically independent variables) where
Gaussian fluctuation of each variable will yield a net deviation from the null hypothesis that is not larger
than the corresponding $\chi^2$ value referenced on the left-hand axis.  The distinction between a one-sided
and two-sided limit is made with regards to the placement of the $0,1,2 \sigma$ bounds relative to the numerical
value of the CDF. Given a substantial excess of events beyond SM expectations, a two-sided limit can in principle be derived for the $\chi^2$ best-fit minimization against $M_{1/2}$, as meaningful deviations may be anticipated in either direction away
from the median.  Particularly small $\chi^2$ values imply a better fit to the data by the assumed
signal (at a certain confidence level) than what could be attributed to random fluctuations around the SM, while excessively
large $\chi^2$ values disfavor a given $M_{1/2}$ relative to the SM-only null hypothesis.  In this case, the usual
$1$ and $2 \sigma$ $68\%$ and $95\%$ consistency integrations enclose areas symmetrically distributed about the mean,
such that centrally inclusive boundary lines are drawn at the CDF percentages $2.2\%$, $15.9\%$, $84.1\%$, and $97.7\%$.
It should be noted that those displayed searches that are statistically dependent are weighted so as to be statistically uncorrelated, specified by the reduced number of degrees of freedom (DOF). A compensating reduction in the effective DOF from the number of actual DOF has the effect of marginally lowering the quoted CDF scale values relative to the left-hand $\chi^2$ axis, slightly compressing the displayed error margins.

The landscape of 7 \& 8 TeV multijet searches with the incidence of a single lepton allow for the isolation of best fits to the rate of \fsu5 SUSY production, as regulated by the mass parameter $M_{1/2}$. The five minima in Figure (\ref{fig:chi_square}) are in excellent agreement, falling within a narrow range of $M_{1/2} = 675$ GeV to $M_{1/2} = 829$ GeV. The $2 \sigma$ $\chi^2$ intersection for all searches in Figure (\ref{fig:chi_square}) sets a lower bound around $M_{1/2} =  680$ GeV.  The depth of the $\chi^2$ well is satisfactory, with the SM limit rising to or above the median probability in four of the five panels, even with compensation for the reduction in effective degrees of freedom. However, it is the closely correlated SUSY fits of all searches that we must emphasize.

Once the gluino-mediated light stop production transitions to on-shell at $M_{1/2} = 729$ GeV, the branching ratio (BR) for $\widetilde{g} \to \widetilde{t}_1 t$ reaches 100\% for all No-Scale \fsu5 points $M_{1/2} \ge 729$ GeV. The light stop may then decay through either $\widetilde{t}_1 \to t \widetilde{\chi}_1^0$ or $\widetilde{t}_1 \to b \widetilde{\chi}_1^{\pm}$, with the $t \widetilde{\chi}_1^0$ state claiming a BR of about 70\% at $M_{1/2} = 750$ GeV, slightly elevated up to a BR of about 76\% at $M_{1/2} = 900$ GeV. Consequently, we can anticipate equivalent final states for gluino, squark, and light stop pair-production at $M_{1/2} \ge 729$ GeV. This equivalency manifests in our data simulation through inclusion of all gluino, squark, and light stop pair-production channels, and not just the light stop pair-production channels that are the central aim of the ATLAS and CMS Refs.~\cite{ATLAS-CONF-2012-073,ATLAS-CONF-2012-166,CMS-PAS-SUS-12-023}. The light stop searches of Refs.~\cite{ATLAS-CONF-2012-073,ATLAS-CONF-2012-166,CMS-PAS-SUS-12-023} do capture more light stop pair-production events by approximately a factor of 10 over the multijet+lepton searches~\cite{ATLAS-CONF-2012-140,Aad:2012naa} that specifically target gluino pair-production, though the overall percentage of light stop pair-production events remains quite low at less than 4\% of all No-Scale \fsu5 simulated events passing all data cuts. The specific percentages for the $\widetilde{t}_1 \widetilde{t}_1$, $\widetilde{g} \widetilde{g}$, $\widetilde{g} \widetilde{q}$, and $\widetilde{q} \widetilde{q}$ channels from our No-Scale \fsu5 Monte-Carlo collider detector simulation are tabulated in Table (\ref{tab:percentages}) for $M_{1/2} = 750$ GeV.

Perhaps the combined effort from all light stop, gluino, and squark pair-production leading to a unified final state reveals the rationale for the such well correlated results between statistically independent SUSY search strategies we have presented here. In order to consistently generate the sufficient number of events to match the excess production above background expectations for all searches in unison, there may have to be a realistic model capable of separate pair-production channels that can lead to a common final state, as exemplified in No-Scale \fsu5, where all light stop, gluino, and squark pair-production channels lead to a multijet final state, many of these events containing a lepton. The CMSSM/mSUGRA and simplified model structures implemented by the LHC Collaborations appear incapable of generating the required productive channels necessary to not just match event production, but also simultaneously correlate SUSY mass scales amongst all statistically independent searches (not to even mention correlating the exquisitely rich Primordial Synthesis of phenomenology to be presented later in this work). To complete a realistic assessment of the SUSY search status at the LHC, the evaluation must gauge the data observations against a truly natural model capable of physically producing the entire data profile that the beam collisions are providing. Thus, we find the contributions of our \fsu5 analysis critical to a comprehensive realistic assessment of the current status in the search for supersymmetry at the LHC.

If in fact the excess production above the background estimates discussed here is attributable to new physics and
not the result of several parallel background fluctuations, then the growth in the signal for larger statistics should remain reasonably
proportional to the increase in luminosity and thus a large deviation of events beyond SM expectations should be clearly evident in the early data acquisition of the 13 TeV LHC scheduled to energize in 2015. It remains to be seen whether these small discrepancies shall ultimately be attributed to primarily statistical or systematic origins in either or both of the independently collected data sets. In view of that, we await the 13 TeV LHC that will clarify whether these event excesses beyond expectations are triggered by fluctuations in the observed 9.7 \fb and 13.0 \fb data or possibly the appearance of new physics.

\begin{figure*}[htp]
        \centering
        \includegraphics[width=0.75\textwidth]{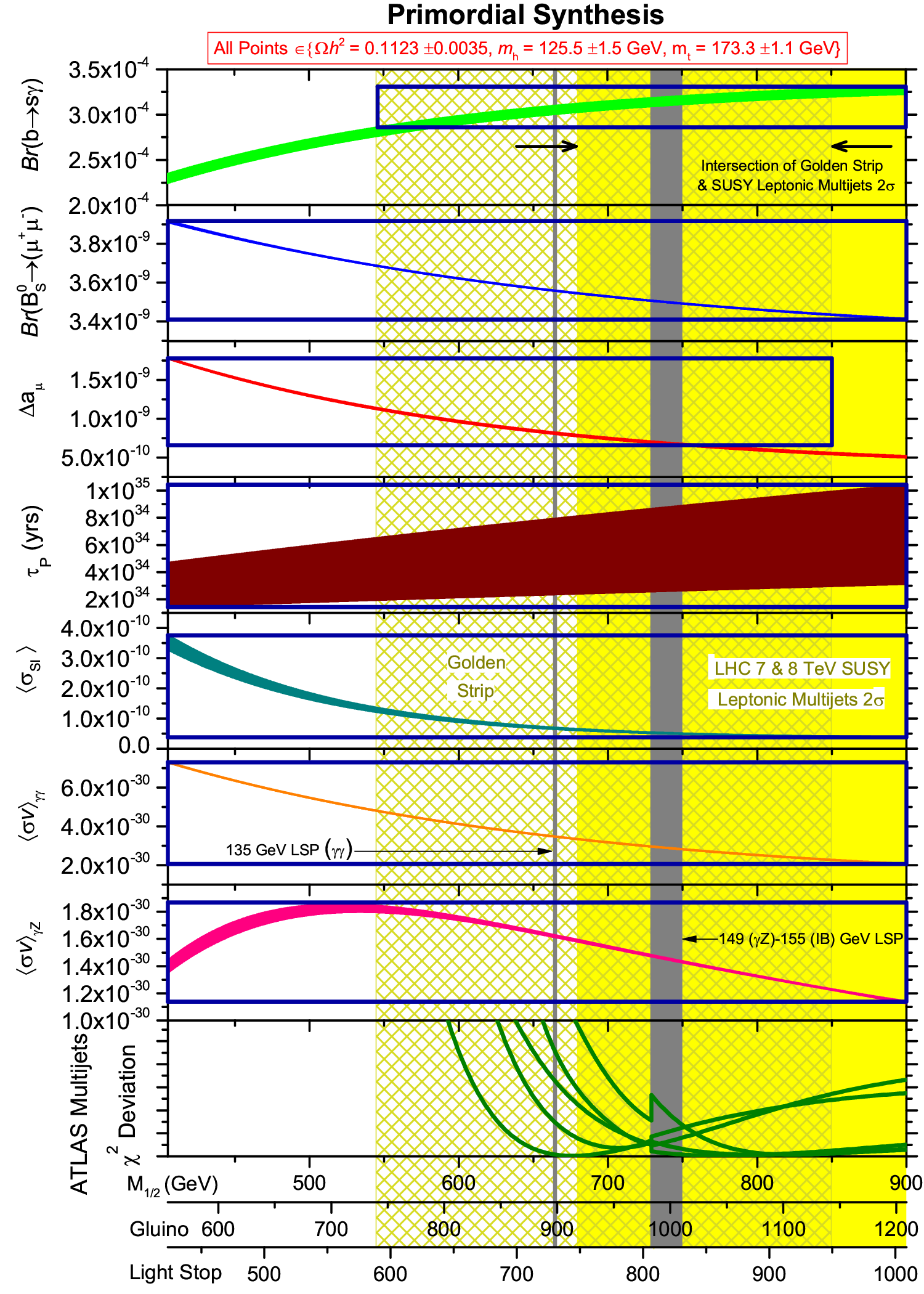}
        \caption{Primordial Synthesis of all currently progressing experiments searching for physics beyond the
Standard Model. All points depicted on each curve satisfy the conditions $0.1088 \le \Omega h^2 \le 0.1158$,
$124 \le m_h \le 127$ GeV, and $172.2 \le m_t \le 174.4$ GeV. Each curve thickness represents an uncertainty on
the strong coupling constant $0.1145 \le \alpha_s(M_Z) \le 0.1172$. The cross-hatched region illustrates the Golden Strip, which identifies the intersection of all current BSM experiments, excluding the LHC collider studies. The constraints imposed upon the BSM experiments shown here are discussed in Section~\ref{sect:pheno}. The yellow region portrays the intersection of all the $\chi^2$ 2$\sigma$ deviations of all the leptonic SUSY searches studied in this work. We highlight the 135 GeV and 149-155 GeV LSP
masses by the vertical strips. Notice that we have inserted the gluino and light stop mass scales at the bottom
of the Figure to enable an instant conversion from the gaugino mass $M_{1/2}$, made possible by the
characteristic rescaling property of the \fsu5 SUSY spectrum in terms of $M_{1/2}$. We only show the model space here up to $M_{1/2} = 900$ GeV, though the yellow band derived from the $\chi^2$ 2$\sigma$ deviation has only the lower bound at $M_{1/2} \ge 680$ GeV. The upper bound on the Golden Strip at $M_{1/2} \le 850$ GeV is solely derived from the lower 2$\sigma$ limit on the ${\rm (g_{\mu}-2)/2}$ measurements.}
        \label{fig:primordial}
\end{figure*}

\section{Primordial Synthesis 2.0}

In light of the new results computed here, we update our Primordial Synthesis of experimental results in Figure (\ref{fig:primordial}), extending the work of Ref.~\cite{Li:2012mr}. For the collider studies in the bottom pane of Figure (\ref{fig:primordial}), in this new version we only include the $\chi^2$ results from the leptonic SUSY searches examined in this work. For all the other BSM experiments in the remaining panes in Figure (\ref{fig:primordial}), we carry forward the exact same limits described and referenced in detail in Ref.~\cite{Li:2012mr}, with the exception of the updated limits for ${\rm (g_{\mu}-2)/2}$ and \textit{Br}(\bs0) as discussed in Section~\ref{sect:pheno}. We have dubbed the non-trivial intersection of all BSM experiments as the Golden Strip~\cite{Li:2010mi,Li:2011xu,Li:2011xg}, as labeled in Figure (\ref{fig:primordial}). We retain identifying strips for 135 GeV and 149-155 GeV LSP masses in Figure (\ref{fig:primordial}), raised slightly from the original 130 GeV and 145-150 GeV LSP masses shown in the corresponding Figure of Ref.~\cite{Li:2012mr}. This small shift is motivated by recent results presented by the FERMI-LAT Collaboration stating that the observed 130 GeV monochromatic gamma-ray line~\cite{Bringmann:2012vr,Weniger:2012tx} is now closer to 135 GeV after data reprocessing~\cite{AAlbert}. This in effect also shifts the $\chi \chi \to \gamma Z$ line to about 149 GeV and internal bremsstrahlung to an estimated 155 GeV, thus we insert a strip for an approximate range of 149-155 GeV to serve as a reference point for LSP masses in \fsu5. It is significant to underline that all the \fsu5 points rendered in Figure (\ref{fig:primordial}) themselves all satisfy the 7-year WMAP constraints $\Omega h^2 = 0.1123 \pm 0.0035$~\cite{Komatsu:2010fb}, the Higgs boson mass $m_h = 125.5 \pm 1.5$ GeV~\cite{:2012gk,:2012gu,Aaltonen:2012qt}, and top quark mass $m_t = 173.3 \pm 1.1$ GeV~\cite{:1900yx}, a very difficult maneuver in and of itself.

The yellow band in Figure (\ref{fig:primordial}) represents the 2$\sigma$ deviation of all the $\chi^2$ curves depicted in Figure (\ref{fig:chi_square}), which has no upper limit and only the lower limit at $M_{1/2} \ge 680$ GeV. The upper limit of the Golden Strip occurs at $M_{1/2} \ge 850$ GeV, where this higher bound is singularly determined by the current measurements of the anomalous magnetic moment ${\rm (g_{\mu}-2)/2}$ of the muon, which is subject to change in the future or even vanishing completely. Nonetheless, we find $any$ intersection between the combination of all LHC SUSY searches with the other BSM experiments to be a non-trivial occurrence, due to the highly constrained nature of both the Golden Strip and LHC constraints. If at some point in the future the lower boundary for $\Delta a_{\mu}$ approaches zero or vanishes entirely, the the upper limit we show here will increase or vanish altogether as well, absent of any upper limit that could possibly be imposed in the future from the LHC SUSY search data.

\section{20 Input Measurements}\label{sect:inputs}

\begin{figure*}[htp]
        \centering
        \includegraphics[width=1.00\textwidth]{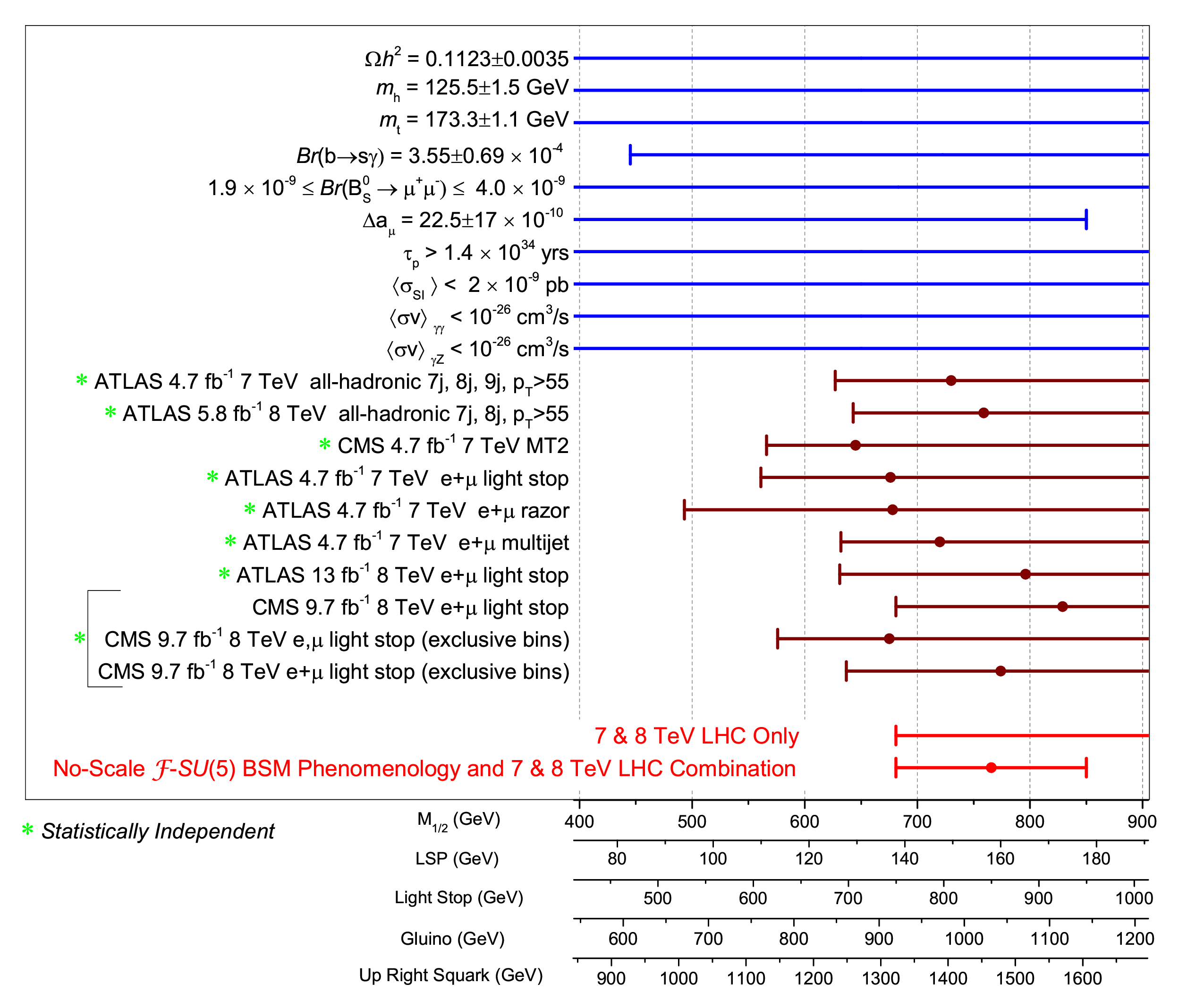}
        \caption{Summarization of all BSM+LHC experiments that are accessible to the No-Scale \fsu5 model space. The limits on the BSM experiments are either 1$\sigma$ or 2$\sigma$, depending on the running time and our confidence in the precision of the result. The constraints imposed upon the BSM experiments shown here are discussed in Section~\ref{sect:pheno}. The ranges on the LHC SUSY searches are all 2$\sigma$ as fit to the No-Scale \fsu5 model, with the dot representing the minimum of the $\chi^2$ curve, or best fit. If no endcap is shown for a particular entry, then the derived constraint either has no limit for that entry or the limit extends beyond the shown mass scale. The green asterisk identifies those SUSY searches that are statistically independent, which comprises eight of the ten SUSY searches included here. We bracket the SUSY searches that share events, even though these do present different perspectives that provide unique insight into the data. The bottom two entries combine all the LHC SUSY searches into a single lower bound, as well as combining this LHC lower bound with the upper bound extracted from the current limits on the BSM experiments, providing a tentative range of $680 \le M_{1/2} \le 850$ GeV. A description of the explicit signal regions depicted here can be found in Section~\ref{sect:susy} for the leptonic searches and Section~\ref{sect:inputs} for the all-hadronic searches. The characteristic rescaling property of No-Scale \fsu5 allows us to conveniently insert SUSY mass scales for relevant sparticles in order to quickly convert from gaugino mass $M_{1/2}$ to spectrum.}
        \label{fig:20inputs}
\end{figure*}

The entire landscape of phenomenological experiments, including LHC, is summarized in Figure (\ref{fig:20inputs}), as it constrains No-Scale \fsu5 in totality. All the BSM experiments shown in Figure (\ref{fig:primordial}) are also exhibited in Figure (\ref{fig:20inputs}), adhering to the same constraints. On top of the leptonic SUSY searches studied in this work, we also incorporate into Figure (\ref{fig:20inputs}) previous \fsu5 LHC investigations~\cite{Li:2012tr,Li:2012ix,Li:2012mr}, updated here inclusive of up to NLO contributions, whereas the previous studies were LO only.

Regarding our earlier statement that the all-hadronic $p_T$ cut on jets should be capped at 55 GeV in order to minimize suppression of possible soft hadronic jets in \fsu5, we implement only the all-hadronic SUSY searches with a $p_T > 55$ GeV cut in Figure (\ref{fig:20inputs}). We also attempt to isolate all events in the $p_T >55$ GeV search regions to establish firm statistical independence in the all-hadronic $\chi^2$, thus embracing search regions of seven jets only and eight jets only. For the search 9j55 that includes all events with at least nine jets, we can isolate the region to only nine jets in the 7 TeV Ref.~\cite{ATLAS-CONF-2012-037}, however, this cannot be likewise accomplished in the 8 TeV Ref.~\cite{ATLAS-CONF-2012-103} since the data for more than nine jets cannot be segregated from nine jets only. This is significant as a result of the large cut on jets of $p_T > 55$ GeV, where we are quite concerned about the region of $\ge10$ jets when implementing such a high cut on jet $p_T$ that could suppress the entire $\ge9$ jet region. We thus resign ourselves here to only the seven jet and eight jet regions for the 5.8 \fb 8 TeV data.

The No-Scale \fsu5 analysis~\cite{Li:2012ix} based upon the CMS 7 TeV $M_{T2}$ variable is updated here to include NLO as well, as shown in Figure (\ref{fig:20inputs}). We simplify the $M_{T2}$ search region by combining the complete low $H_T$ and high $H_T$ regions of Ref.~\cite{Chatrchyan:2012jx}, now encompassing only a single large search region of $H_T \ge 750$ GeV  and $M_{T2} \ge 150$ GeV. Again, striving for statistically independent LHC SUSY searches in Figure (\ref{fig:20inputs}), we omit the $M_{T2} b$ strategy of Ref.~\cite{Chatrchyan:2012jx} as a result of it being a wholly contained subset of the $M_{T2}$ data set, and thus statistically dependent with the $M_{T2}$ search region.

Through these generalizations, Figure (\ref{fig:20inputs}) now involves all statistically independent LHC SUSY search results, with the exception of our exclusive binning tactic for the CMS 8 TeV leptonic light stop search described earlier in this work, though we reiterate that the exclusive binning is mandatory due to its ability to present the data in a unique format that offers a completely different perspective than just the entire search region. The 8 TeV statistics are based upon the 2012 collisions only, and do not include the 2010-11 7 TeV collisions, representing 100\% statistically independent results, as obviously are the ATLAS and CMS results themselves. In Figure (\ref{fig:20inputs}) we show the lower 2$\sigma$ limits on the LHC SUSY searches, identified by the range for each entry. The dot annotates the minimum of the $\chi^2$ well obtained in Figure (\ref{fig:chi_square}). We focus in Figure (\ref{fig:20inputs}) on only those LHC SUSY searches that are sensitive and hence probing of the No-Scale \fsu5 parameter space, namely multijet events. At this juncture, Figure (\ref{fig:20inputs}) now consists of eight statistically independent SUSY search results, identified by a green asterisk, all accessible to the \fsu5 model space. 

The impressive attribute of Figure (\ref{fig:20inputs}) is the correlation currently being experienced by all BSM+LHC experiments in No-Scale \fsu5, derived from 20 input measurements. The intersection of all 20 BSM+LHC experimental measurements provide us a global fit of $M_{1/2}$ = 680--850 GeV, corresponding to sparticle masses of $M(\widetilde{\chi}_1^0)$ = 138--178 GeV, $M(\widetilde{t}_1)$ = 750--950 GeV, $M(\widetilde{g})$ = 920--1150 GeV, and $M(\widetilde{u}_R)$ = 1320--1600 GeV. The central value of this span falls at $M_{1/2} = 765$ GeV, with $M(\widetilde{\chi}_1^0)$ = 158 GeV, $M(\widetilde{t}_1)$ = 850 GeV, $M(\widetilde{g})$ = 1030 GeV, and $M(\widetilde{u}_R)$ = 1460 GeV. To reiterate, the lower bounds here are derived from the LHC SUSY search lower limits, while the upper bounds result from the lower 2$\sigma$ limit on the latest ${\rm (g_{\mu}-2)/2}$ measurements.

These sparticle mass fits distinctly derived for No-Scale \fsu5 may or may not conflict with certain simplified model limits established by the LHC Collaborations in their 8 TeV results, however, we caution against an overly literal application of the mass limits established via these Collaboration surveys to the \fsu5 context. Such examples often decouple all SUSY fields other than the neutralino and gluino octet, or otherwise impose strict mass degeneracies, and may unrealistically force key branching ratios to unity. We suggest that studies such as our own here may fill a vital gap of information by providing independent data-driven demarcation of mass limits specifically applicable to certain realistic models, such as No-Scale \fsu5.

\section{Conclusions}

We have investigated the 7 TeV and 8 TeV ATLAS and CMS SUSY searches requiring one lepton final state in the framework of the supersymmetric GUT model No-Scale Flipped $SU(5)$ with additional vector-like flippon multiplets, or \fsu5 for brevity. Only those strategies implementing a low $p_T$ cut on hadronic jets have been incorporated into our analysis as a result of the possibility of soft hadronic jets if the SUSY mass scale lives near an off-shell to on-shell transition for gluino-mediated light stop production. This encompassed three 4.7 \fb ATLAS 7 TeV gluino and light stop searches, along with 13 \fb ATLAS 8 TeV and 9.7 \fb CMS 8 TeV light stop searches. This group of SUSY searches represent five statistically independent SUSY multijet search strategies applying $p_T > 20-30$ GeV, close to our originally motivated \fsu5 multijet SUSY discovery vision of $p_T > 20$ GeV necessary to thoroughly probe the No-Scale \fsu5 viable parameter space.

First, we discussed how an analysis of ATLAS and CMS SUSY search results in the context of only the peak in the ${\rm E_T^{Miss}}$ distribution could be undertaken, if data warrants, in order to uncover visible signs of No-Scale \fsu5 SUSY, with zero input from Monte Carlo statistics. We had previously shown that in No-Scale \fsu5, the peak in the ${\rm E_T^{Miss}}$ distribution occurs at about twice the LSP mass, permitting a quick translation of the ${\rm E_T^{Miss}}$ peak to the LSP mass in \fsu5 and hence to the SUSY mass scale $M_{1/2}$ and the sparticle spectrum.

Additionally, we introduced a statistical isolation method to systematically quantify the pairwise overlap between any two basis elements of the
compound search space. This provides a concrete method for the statistical disentanglement of correlated searches, and a quantitative measure of statistical independence of LHC SUSY searches.

We next executed a full Monte-Carlo analysis, and our results conclude that the $best~fits$ of the $\chi^2$ of all five mutually exclusive leptonic data sets fall within an \fsu5 SUSY mass scale ranging from $M_{1/2} = 675$ GeV to $M_{1/2} = 829$ GeV, displaying fine consistency from 7 TeV to 8 TeV and also amongst the two LHC Collaborations. Our Monte-Carlo procedure includes all QCD contributions up to next-to-leading order in the strong coupling constant, exhibiting agreement to around $\sim 20$\% with the Collaboration Monte-Carlo results for common simplified benchmarks after all selection cuts, accounting for our ommission of resummation of soft gluon emission at next-to-leading-logarithmic accuracy (NLO+NLL). 

At a more fundamental level, we further observe the required stability between all the leptonic and all-hadronic collider studies sensitive to the \fsu5 model space and all parallel BSM experiments, where an intersection amongst all experiments (non--LHC + LHC) condenses to a SUSY mass scale of $M_{1/2} = 680-850$ GeV. This range for the gaugino mass $M_{1/2}$ corresponds to sparticle masses of $M(\widetilde{\chi}_1^0)$ = 138--178 GeV, $M(\widetilde{t}_1)$ = 750--950 GeV, $M(\widetilde{g})$ = 920--1150 GeV, and $M(\widetilde{u}_R)$ = 1320--1600 GeV. We emphasized that the upper bound at $M_{1/2} \le 850$ GeV was solely due to the lower limit on our combination of the most recent $e^+ e^-$ and $\tau^{\pm}$ measurements of the anomalous magnetic moment ${\rm (g_{\mu}-2)/2}$ of the muon, while the lower bound at $M_{1/2} \ge 680$ GeV results from the lower 2$\sigma$ limit on the ATLAS and CMS LHC SUSY searches analyzed here in the context of a No-Scale \fsu5 framework. If at some point the lower limit for $\Delta a_{\mu}$ approaches zero or vanishes completely, then the upper limit on $M_{1/2}$ we depict here will increase or vanish altogether as well, absent of any upper bound that could potentially be imposed in the future from LHC SUSY search data.

Most importantly, we look ahead to continued constancy in the best SUSY mass fit of the LHC observations within the framework of \fsu5. We have already witnessed low variability in the best SUSY mass fit in \fsu5 from 4.7 \fb at 7 TeV to 9.7 \fb and 13 \fb at 8 TeV. If the leptonic SUSY searches at the LHC still cannot exclude the SUSY mass scale range at $M_{1/2} = 680-850$ GeV post conclusion of the 8 TeV 20 \fb results, then the high-energy physics community could have much to anticipate when the 13 TeV LHC launches in 2015.


\begin{acknowledgments}
This research was
supported in part
by the DOE grant DE-FG03-95-Er-40917 (TL and DVN),
by the Natural Science Foundation of China
under grant numbers 10821504, 11075194, 11135003, and 11275246 (TL),
and by the Mitchell-Heep Chair in High Energy Physics (JAM).
We also thank Sam Houston State University
for providing high performance computing resources.
\end{acknowledgments}


\bibliography{bibliography}

\end{document}